\documentclass[a4paper,11pt]{article}
\usepackage{pos}

\title{Constraining the quark and gluon helicity at STAR}

\author{Ting Lin for the STAR Collaboration}

\affiliation{Institute of Frontier and Interdisciplinary Science \& Key Laboratory of Particle Physics and Particle Irradiation (MoE), Shandong University, Qingdao, Shandong 266237, China}

\emailAdd{tinglin@sdu.edu.cn}

\abstract{
        How quarks and gluons conspire to provide the total spin of proton is a long-standing puzzle in quantum chromodynamics (QCD). The unique capability of RHIC, that can provide longitudinally polarized $p+p$ collisions at both $\sqrt{s} = 200$~GeV and $\sqrt{s} = 510$~GeV, opened new territory to constrain the helicity structure of the proton with unprecedented depth and precision.\par

        Results from various STAR spin measurements have contributed significantly to our understanding of the quark and gluon helicity distributions inside the proton. The longitudinal double-spin asymmetry, $A_{LL}$, from the STAR 2009 inclusive jet measurement, provides the first indication of the positive gluon polarization with partonic momentum fraction $x$ greater than 0.05 inside the proton. More precise measurements using the $p+p$ data collected in 2012, 2013 and 2015 at both $\sqrt{s}$ = 510 and 200 GeV confirm the previous findings and provide additional constraints in the largely unexplored region of $x < 0.05$. Compared to the inclusive jet observables, analyses of dijet production extending to higher pseudorapidity (up to $\eta \sim 1.8$) provide better constraints on the $x$ dependent behavior of $\Delta g(x)$. Moreover, the reconstruction of $W^{\pm}$ in longitudinally polarized proton-proton collisions provides significant constraints on the flavor separation of the light sea quark helicity distributions inside the proton, while the longitudinal spin transfer to $\Lambda$ and $\bar \Lambda$ hyperons provides access to the helicity of strange and anti-strange quarks in the proton.\par

        In this proceeding, an overview of dijet measurements from longitudinally polarized proton-proton collisions at STAR is presented.
}

\FullConference{25th International Spin Physics Symposium (SPIN 2023)\\
 24-29 September 2023\\
 Durham, NC, USA\\}

\begin{document}
\maketitle

\section{Introduction}

Understanding how the quark and gluon's intrinsic spins and their orbital angular momenta combine to make up the proton spin of $\hbar/2$ has been a major challenge in quantum chromodynamics (QCD)~\cite{Jaffe:1989jz}. Longitudinally polarized proton-proton collisions at the Relativistic Heavy Ion Collider (RHIC)~\cite{Alekseev:2003sk} serve as a crucial experimental avenue to unravel the intricate spin structure of the proton with various final state interactions in the experiment. In these collisions, protons are polarized along the direction of their momentum, providing a unique opportunity to probe the individual components contributing to the spin of the proton.\par

The Solenoidal Tracker at RHIC (STAR) is a large acceptance detector located at RHIC facility, with over two decades of continuous and successful operation. The central part of the STAR detector is the Time Projection Chamber (TPC), that measures the momentum of charged particles scattered within $|\eta| < 1.3$ in a 0.5~T solenoidal magnetic field. Barrel and Endcap Electromagnetic Calorimeters (BEMC and EEMC) serve dual roles, not only measuring electromagnetic energy depositions at $-1.0 < \eta < 2.0$ but also operating as triggering detectors in polarized $p+p$ collisions. The vertex position detector (VPD) and the zero degree calorimeters (ZDC) are pairs of far-forward detectors that provide the measurement of the relative luminosities of the colliding bunches associated with a given helicity state~\cite{STAR:2002eio}.\par

RHIC conducts longitudinally polarized proton-proton collisions at both $\sqrt{s} =$ 200~GeV and 510~GeV. In 2009 and 2015, STAR recorded integrated luminosities of 25 $\mathrm{pb^{-1}}$ and 52 $\mathrm{pb^{-1}}$ at $\sqrt{s} = 200$ GeV, with average beam polarizations of approximately $55\%$ and $58\%$, respectively. While in 2012 and 2013, a substantial dataset was acquired at $\sqrt{s} = 510$ GeV. The integrated luminosity reached approximately 82 $\mathrm{pb^{-1}}$ in 2012, with an average beam polarization about $53\%$, and then increased to 300~$\mathrm{pb^{-1}}$ in 2013 with about $55\%$ average polarization.\par

In hadronic interactions, the hard scattering quarks or gluons undergo a process called fragmentation and hadronization due to the non-perturbative nature of QCD. This process results in a creation of a collimated spray of particles, primarily hadrons, collectively referred to as a jet. Jets serve as crucial signatures in the study of hadronic interactions, offering valuable insights into the dynamics of underlying quarks and gluons. STAR pioneered the novel measurements of jet, dijet and jet substructure to probe the internal spin structure of the proton. High precision measurements of jet and dijet longitudinal double spin asymmetries from STAR have revealed, for the first time, that there is a sizable polarization of the gluon inside the proton at momentum fractions $x > 0.05$ \cite{deFlorian:2014yva, Nocera:2014gqa, Adamczyk:2015}.\par 

\section{Gluon Polarization Measurement in Dijet Production}

In longitudinally polarized proton-proton collisions, the measured observable is the longitudinal double spin asymmetry ($A_{LL}$). It is the ratio of the longitudinally polarized cross section over the unpolarized one. Experimentally, sorting the measured yields by beam spin state, and combining other independent measurements, the asymmetry is evaluated as:
\begin{equation}
	A_{LL} = \frac{\sigma^{++} - \sigma^{+-}}{\sigma^{++} + \sigma^{+-}} = \frac{\sum P_{Y}P_{B}(N^{++} - rN^{+-})}{\sum P_{Y}^{2}P_{B}^{2}(N^{++} + rN^{+-})} 
\end{equation}
where the ratio of the differential cross section ($\sigma$) is further converted into the ratio of jet or dijet yields ($N$), which is then scaled by the luminosity ratio ($r$) for same ($++$) and opposite ($+-$) helicity beams. Here, $P_{Y}$ and $P_{B}$ denote the polarizations of the yellow and blue beams, respectively. $A_{LL}$ is approximately proportional to the product of the helicities ($\Delta f$) of the two interacting partons over the unpolarized parton distribution functions ($f$). The partonic asymmetry ($\hat{a}_{LL}$) which is calculable from perturbative QCD and prominently large at leading order, further influences $A_{LL} \sim \frac{\Delta f_{a}\Delta f_{b}}{f_{a}f_{b}}\hat{a}_{LL}$. At RHIC energy, the dominant subprocesses in hard scattering are gluon-gluon and quark-gluon interactions, all of these make the $A_{LL}$ an ideal probe to the gluon polarization inside proton.\par

In contrast to the inclusive jet measurement, extending the measurement to dijet production allows for better constraints on the $x$ dependence of the gluon's helicity distribution. At leading order, the dijet invariant mass is proportional to the square root of the product of the partonic momentum fractions, $M = \sqrt{sx_{1}x_{2}}$. And the sum of the jet pair pseudorapidities determines their ratio, $\eta_{3} + \eta_{4} = \mathrm{ln}(x_{1}/x_{2})$. With different topological configurations for the dijet production, we can systematically explore distinct combinations of the partonic interactions with symmetric ($x_{1} = x_{2}$) and asymmetric ($x_{1} > x_{2}$ or $x_{1} < x_{2}$) momentum fractions.\par

\begin{figure*}
\centering
\begin{minipage}{.49\textwidth}
\vspace*{1.45cm}
    \includegraphics[width=1\columnwidth]{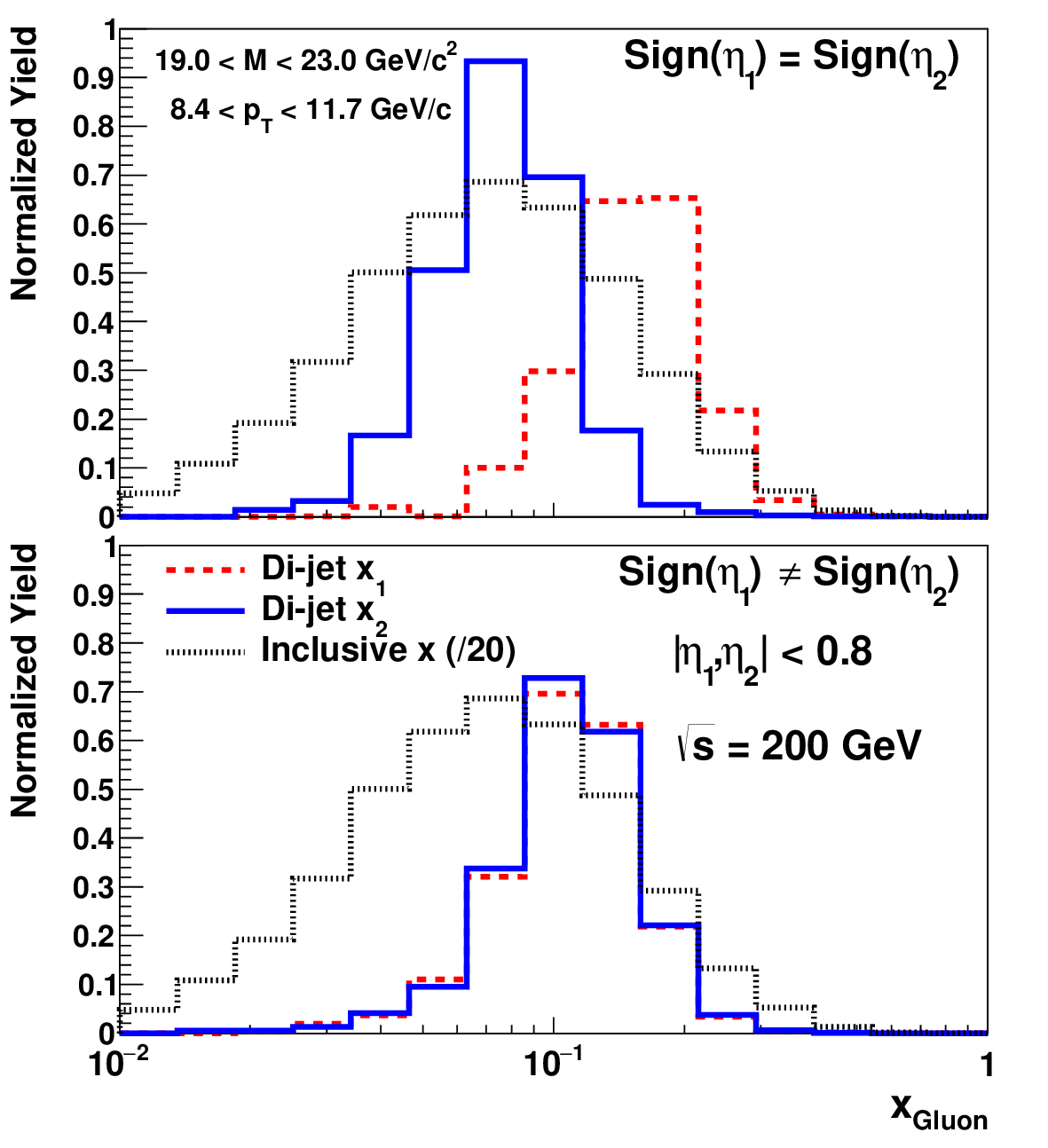}
        \caption{ The distributions of the parton $x_1$ and $x_2$ from \textsc{Pythia} simulations at $\sqrt{s}$ = 200 GeV for two different dijet topologies at mid-rapidity, compared to the gluon $x$ distribution for inclusive jets scaled by an additional factor of $20$ in each panel from \cite{Adamczyk:2016okk}.}
    \label{fig:Dijet_x_frac}
\end{minipage}
\hfill
\begin{minipage}{.49\textwidth}
  \includegraphics[width=1\columnwidth]{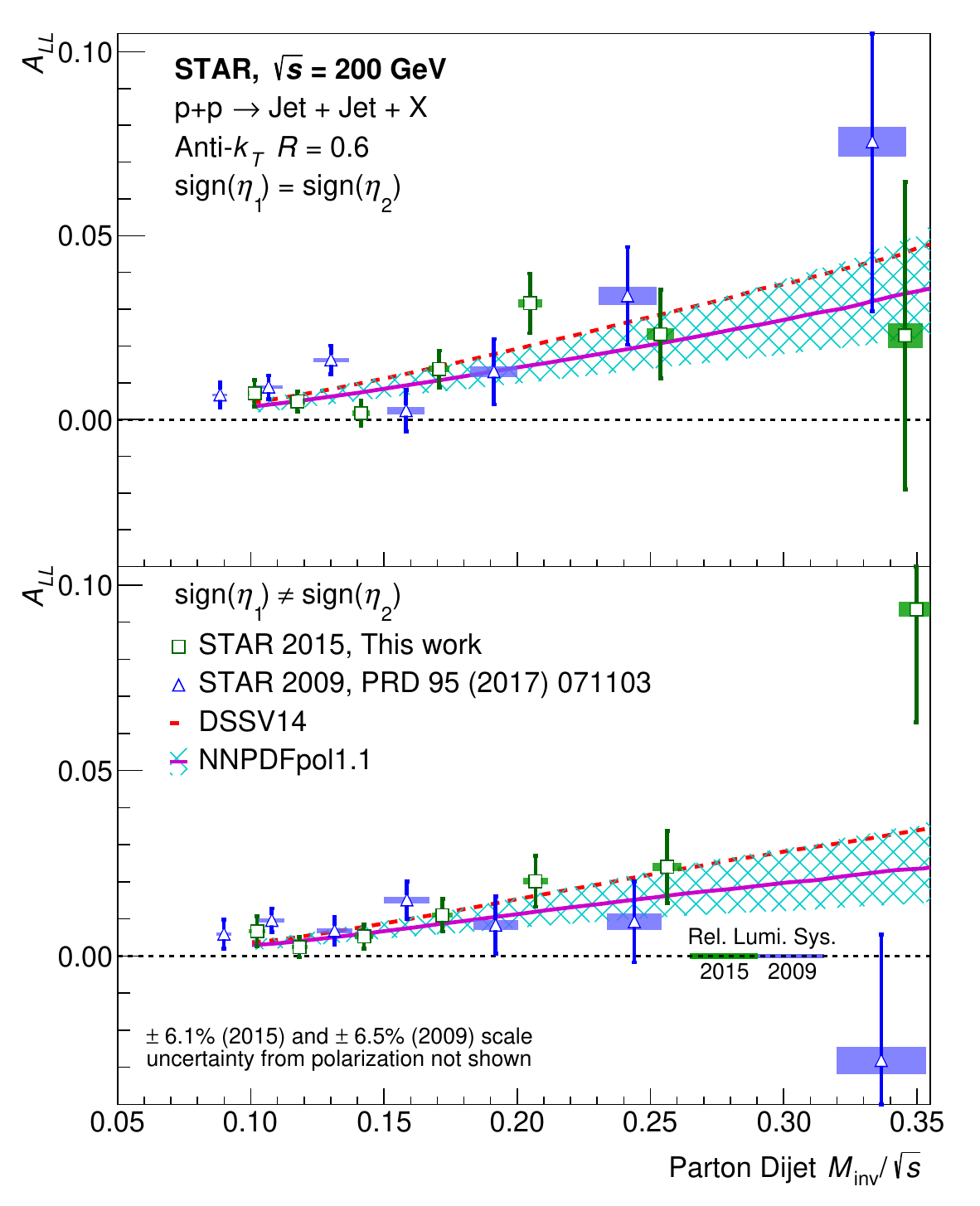}
  \caption{$A_{LL}$ as a function of parton-level dijet invariant mass for dijets with the sign($\eta_{1}$) $=$ sign($\eta_{2}$) (top) and sign($\eta_{1}$) $\neq$ sign($\eta_{2}$) (bottom) event topologies from~\cite{PhysRevD.103.L091103}.}
  \label{fig:Dijet_ALL_pp200}
\end{minipage}
\end{figure*}

Figure~\ref{fig:Dijet_x_frac} illustrates the partonic momentum fraction distributions of the gluons from PYTHIA for dijets with invariant mass of $19.0 < M < 23.0$ GeV$/c^{2}$ compared with momentum fraction probed by inclusive jets at $8.4 < p_{T} < 11.7$ GeV$/c$~\cite{Adamczyk:2016okk}. These distributions are presented for two distinct topologies at $\sqrt{s} =$ 200 GeV measurement. The top panel shows the “same-sign” configuration, where both jets exhibit either positive or negative pseudorapidity, selecting for more asymmetric collisions. Conversely, the bottom panel shows events stemming from relatively symmetric partonic collisions in the “opposite-sign” configuration, featuring one positive pseudorapidity jet with a negative counterpart. The final asymmetries for dijet $A_{LL}$ from 2009 (blue)~\cite{Adamczyk:2016okk} and 2015 (green)~\cite{PhysRevD.103.L091103} measurements are presented in Fig.~\ref{fig:Dijet_ALL_pp200}. Good agreements are found between these two independent measurements.\par

\begin{figure*}
\centering
\begin{minipage}{.49\textwidth}
    \includegraphics[width=1\columnwidth]{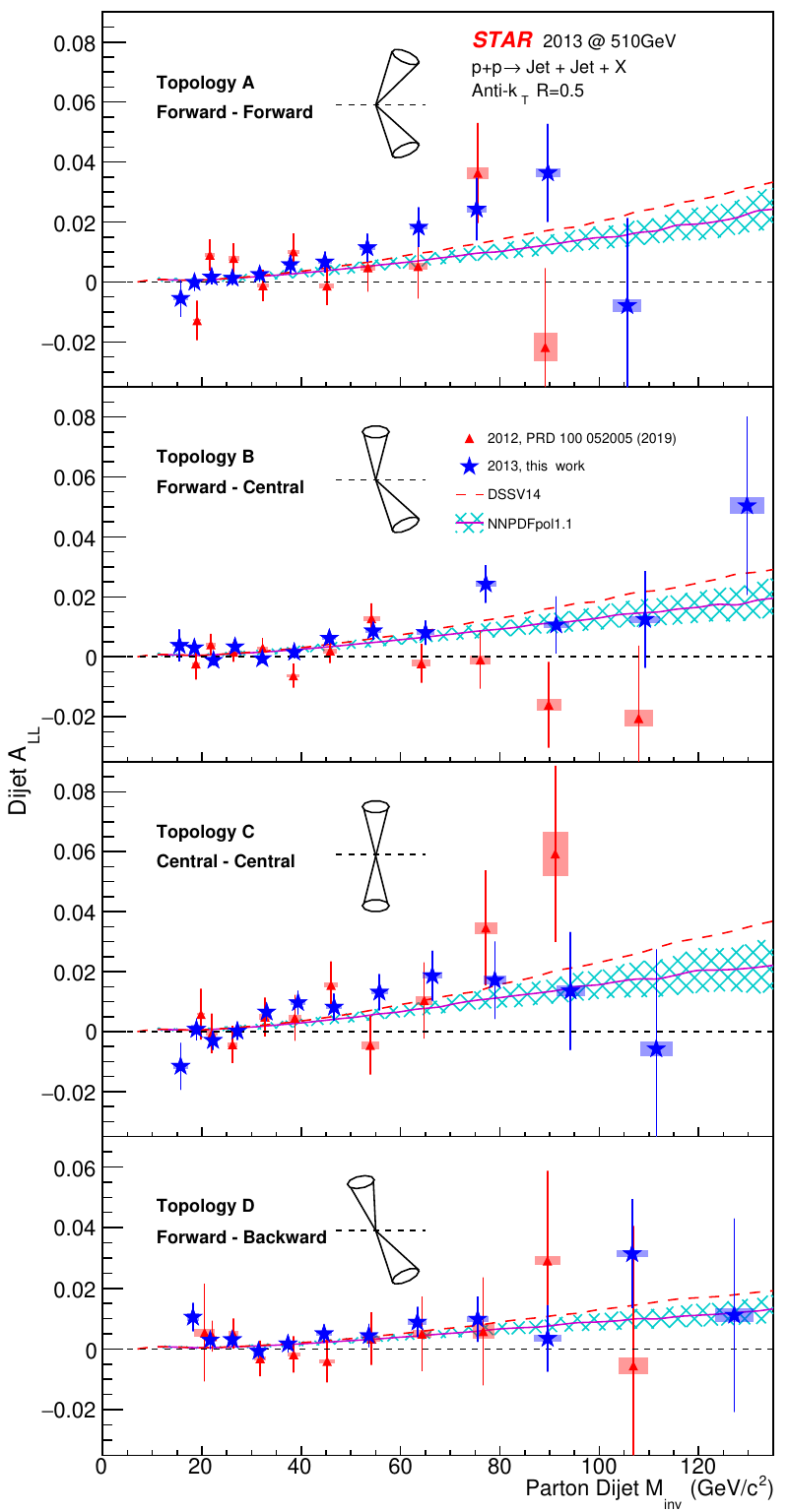}
    \caption{$A_{LL}$ as a function of parton-level dijet invariant mass for dijets with the four different event topologies from~\cite{STAR:2021mqa}.}
    \label{fig:Dijet_ALL_pp510}
\end{minipage}
\hfill
\begin{minipage}{.49\textwidth}
\vspace*{2.4cm} 
  \includegraphics[width=1\columnwidth]{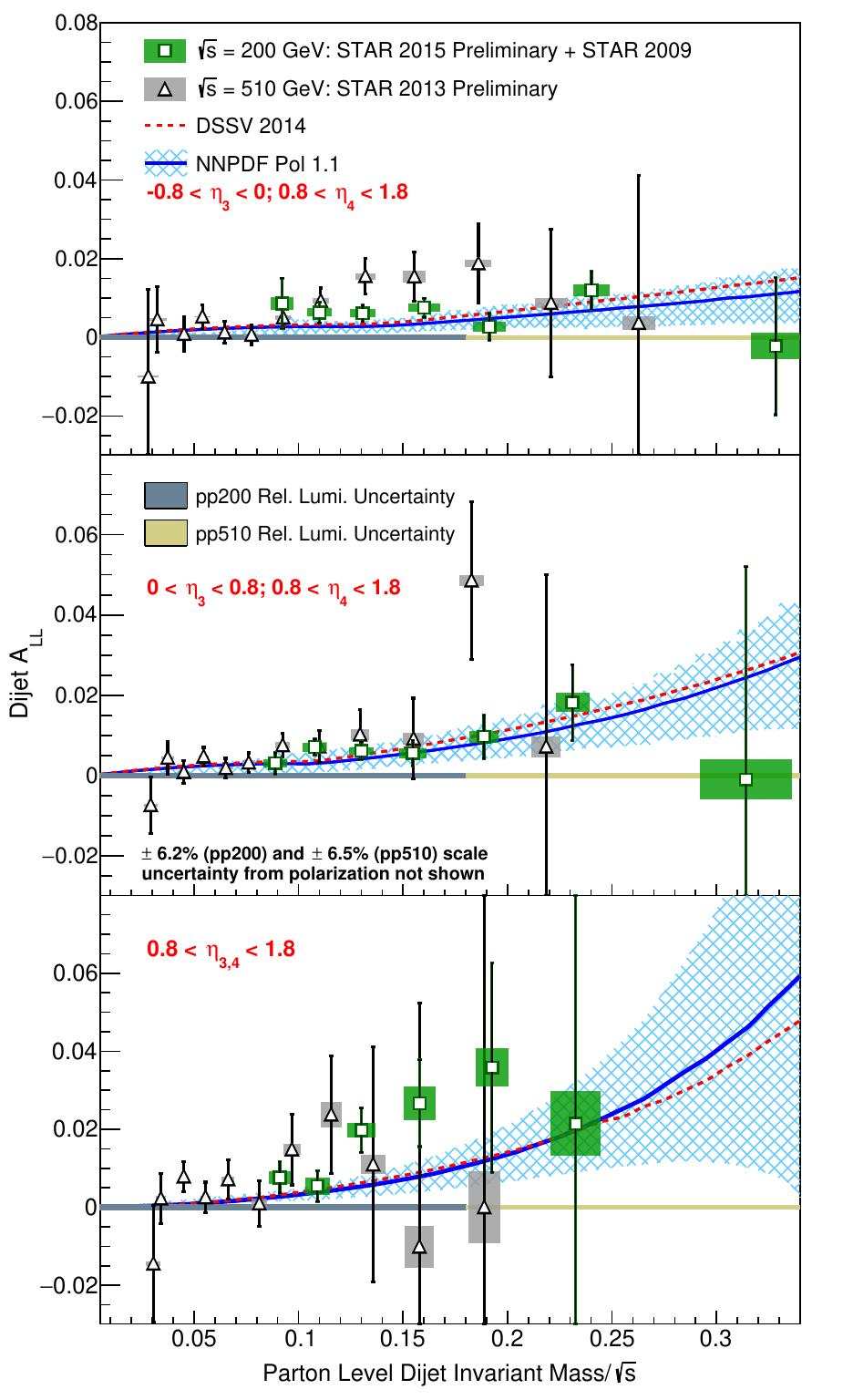}
    \caption{$A_{LL}$ as a function of parton-level dijet invariant mass for dijets with the East barrel-endcap (top), West barrel-endcap (middle) and endcap-endcap (bottom) event topologies.}
  \label{fig:Dijet_ALL_pp200_pp510}
\end{minipage}
\end{figure*}

With the same dijet invariant mass, the increased center-of-mass energy extends the sensitivity to lower momentum fraction partons. Figure~\ref{fig:Dijet_ALL_pp510} shows $A_{LL}$ from the 510 GeV measurement, utilizing the data recorded in 2012 (red)~\cite{PhysRevD.100.052005} and 2013 (blue)~\cite{STAR:2021mqa}. The dijet $A_{LL}$ results are plotted as a function of the parton level invariant mass across four pseudorapidity topologies, facilitating the extraction of $x-$dependent constraints with respect to the dijet invariant mass. In these four configurations, topologies A and B probe the most asymmetric collisions, while C and D originate from more symmetric partonic collisions. Additionally, A and C sample collisions involving partonic interactions with a small scattering angle in the partonic center-of-mass frame where partonic asymmetry $\hat{a}_{LL}$ is expected to be large~\cite{STAR:2021mqa}. B and D topologies roughly sample collisions with a large scattering angle, corresponding to a smaller partonic asymmetry $\hat{a}_{LL}$.\par

With the same center-of-mass energy, we can probe more asymmetric nature of the collisions by extending the measurement into higher pseudorapidity. Figure~\ref{fig:Dijet_ALL_pp200_pp510} shows the $A_{LL}$ results with at least one of the jet detected at higher pseudorapidity in the Endcap ($0.8 < \eta_{\rm jet} < 1.8$) region for both the $\sqrt{s} =$ 200 GeV and 510 GeV measurements. These results are presented as a function of the dijet invariant mass over the collision energy, separated into three distinct dijet topologies with increasing separation on $x_{1}$ and $x_{2}$ from the top to the bottom panels. Good consistency is observed among these two measurements, suggesting a weak collision energy dependence for dijet $A_{LL}$ in hadronic interactions~\cite{PhysRevD.98.032011}.\par

Furthermore, all the dijet results exhibit excellent agreement with theoretical predictions from DSSV~\cite{deFlorian:2014yva} and NNPDF~\cite{Nocera:2014gqa} groups. These predictions, incorporating our prior inclusive jet data, indicate sizable gluon polarization inside proton~\cite{Adamczyk:2015}.\par

The DSSV group has conducted a comprehensive assessment of recent STAR data, including both the inclusive jet and dijet measurements. Employing a Monte Carlo sampling strategy, helicity parton densities were also extracted from STAR dijet results, as outlined in~\cite{DeFlorian:2019xxt}. This approach avoids the need for adopting a tolerance criterion and addresses various shortcomings associated with the propagation of PDF uncertainties to experimental observables. More recently, significant progress has been made with the introduction of the first NNLO corrections for jet production in polarized $p+p$ collisions, aiming to further reduce the theoretical uncertainties. These combined advancements in both experimental and theoretical sides improve the precision of gluon helicity across mid to low momentum fraction regions. As a result, the current estimate for gluon polarization is approximately $0.3 \pm 0.1$ in the $x > 0.01$ region~\cite{WernerVogelsangSPIN2023}.\par

\section{Conclusion}

RHIC concluded its longitudinal polarized data collection in 2015, with nearly two decades of substantial contributions to our comprehensive understanding of the proton's internal spin structure. Results from STAR revealed a sizable asymmetry in the polarized anti-quark sea. Notably, jet and dijet results from STAR indicated, for the first time, a preference for gluon spins to align in the same direction as the proton spin in $x > 0.05$ region. With a few remaining results to be published soon, STAR will conclude the analysis of gluon polarization, a topic that will be revisited in the future Electron-Ion Collider (EIC).

This work was supported in part by the National Natural Science Foundation of China (Grant No.12375139).\par

\end{document}